\documentclass{PHYEAUTH}
\usepackage{graphicx}
\usepackage{amsmath}
\usepackage{amssymb}

\begin{document}

\begin{frontmatter}

\title{Singlet-triplet oscillations and far-infrared spectrum of
four-minima quantum-dot molecule}

\author{M. Marlo-Helle\thanksref{thank1},}
\author{A. Harju, and R. M. Nieminen}

\address{Laboratory of Physics, Helsinki University of Technology,
P.O. Box 1100, FIN-02015 HUT, Finland}

\thanks[thank1]{
Corresponding author. Fax: +358 9 451 2177. \\
 {\it E-mail address}: Meri.Helle@hut.fi}

\begin{abstract}
We study ground states and far-infrared spectra (FIR) of two electrons
in four-minima quantum-dot molecule in magnetic field by exact
diagonalization. Ground states consist of altering singlet and triplet
states, whose frequency, as a function of magnetic field, increases
with increasing dot-dot separation. When the Zeeman energy is
included, only the two first singlet states remain as ground states.
In the FIR spectra, we observe discontinuities due to crossing ground
states. Non-circular symmetry induces anticrossings, and also an
additional mode above $\omega_+$ in the spin-triplet spectrum.  In
particular, we conclude that electron-electron interactions cause only
minor changes to the FIR spectra and deviations from the Kohn modes
result from the low-symmetry confinement potential.
\end{abstract}


\end{frontmatter}


\section{Introduction}
\label{Intro}

Many experimental and theoretical studies have revealed interesting
properties of few-electron quantum dots (QDs) \cite{kirja}. Rich
spectrum of crossing energy levels as a function of magnetic field and
strong interaction effects are nowdays rather well understood in a
symmetric confinement potential.  Recently, the focus has turned into
understanding properties of quantum dots in less symmetric confinement
potentials. In a circular-symmetric confinement potential the center
of mass and relative variables decouple, which makes especially the
excitation spectra trivial \cite{MaksymChakra}. In a less symmetric
confinement this condition is lifted.  However, it is not clear how
the symmetry of the confinement and interaction effects show up in the
far-infrared excitation spectra (FIR) of a low-symmetry QD.

In this work we examine ground states and far-infrared excitation
spectra of two electrons in four-minima confinement potential. Ground
state of four-minima quantum-dot molecule (QDM) consist of altering
spin-singlet ($S=0$) and spin-triplet ($S=1$) states as a function of
magnetic field. On contrary to two-minima QDM \cite{AriPRL}, the
second singlet region in four-minima QDM can be observed at the
greatest inter-dot distances studied even if the Zeeman energy is
included. In FIR spectra we observe anticrossings in Kohn modes and an
additional mode above $\omega_+$ in spin-triplet spectrum. Crossing
ground state levels induce discontinuities in the two-body FIR
spectra. In particular, as in two-minima QDM \cite{MeriPRL}, we
conclude that electron-electron interactions cause only minor changes
to the FIR spectra and deviations from the Kohn modes result in from
the low-symmetry confinement potential.

\section{Model and method}

We model the two-electron QDM by a 2D Hamiltonian
\begin{equation}
H = \sum _{i=1}^2\left ( \frac{ ( {- i {\hbar} \nabla_i}
-\frac ec \mathbf{A})^2 }{2 m^{*}} + V_\mathrm{c}({\bf
r}_{i}) \right ) +  \frac {e^{2}}{ \epsilon   r_{12} } \ ,
\label{ham}
\end{equation}
where $V_\mathrm{c}$ is the external confinement potential:
\begin{equation}
 V_\mathrm{c} = \frac 12 m^* \omega_0^2 \min [(\mathbf{r} + \mathbf{L}_i)^2].
\end{equation}
The potential consists of four parabolas with minima at positions
$\mathbf{L}_i=(\pm L,\pm L)$.  We use the GaAs material parameters
$m^*/m_e=0.067$ and $\epsilon=12.4$, and the confinement strength
$\hbar\omega_0=3.0$~meV.  $\mathbf{A}$ is the vector potential of the
magnetic field (along the $z$ axis) taken in the symmetric gauge. The
Hamiltonian is spin-free and the Zeeman energy can be included in the
total energy afterwards $E_Z = g^*\mu_B B S_Z$ ($g^* = -0.44$ for
GaAs).

We construct two-body wave functions, with total spin $S$,
\begin{eqnarray}
\Psi_S({\mathbf r}_1,{\mathbf r}_2) = \sum_{i\le j} \alpha_{i,j} \{
\phi_i (\mathbf{r}_1) \phi_j(\mathbf{r}_2) + \nonumber \\ (-1)^S\phi_j
(\mathbf{r}_1) \phi_i(\mathbf{r}_2) \},
\end{eqnarray}
using 2D gaussians ($\phi_i(\mathbf{r}) = x^{n_{x_i}} y^{n_{y_i}}
 \e^{-\frac 12 r^2}$) as a single particle basis. (See
Ref. \cite{AriPRL} for more details). The Hamiltonian matrix is
diagonalized numerically.

In the calculation of far-infrared spectra we use Fermi golden rule
within electric-dipole approximation to calculate the transition
probability from the ground state ($E_0$) to excited states ($E_l$):
\begin{equation}
\mathcal{A}_{l, \pm} \propto \left| \left< \Psi_{l} \left| e^{\pm i
\phi} \textrm{$\sum _{i=1}^{2}$} \mathbf{r}_i \right| \Psi_{0} \right>
\right| ^2 \delta(E_l-E_0-\hbar \omega) \ ,
\end{equation}
where $\pm$ refers to two circular polarizations.

\section{Singlet-triplet oscillations}

\begin{figure}
\begin{center}
\includegraphics*[width=0.8\columnwidth]{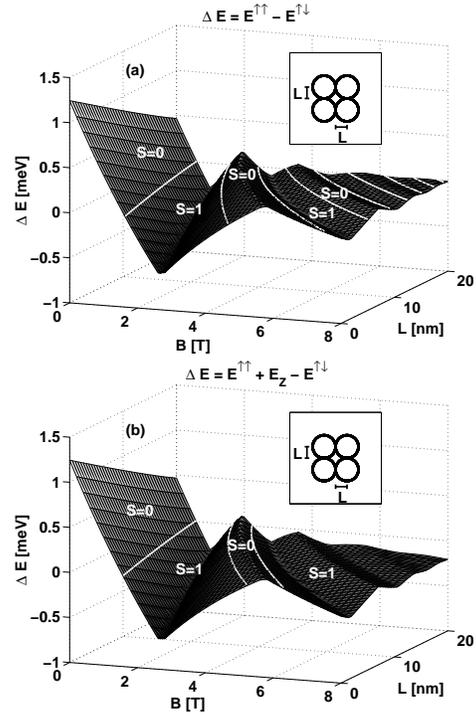}
\caption{Triplet singlet energy difference ($\Delta E = E^{\uparrow
\uparrow} - E^{\uparrow \downarrow}$) in two-electron four-minima
quantum dot molecule as a function of magnetic field and dot-dot
separation $L$. In (a) the energy difference is plotted without Zeeman
energy and in (b) with Zeeman energy. Insets show geometry of
four-minima QDM.}
\label{dE}
\end{center}
\end{figure}

\begin{figure*}
\begin{center}
\includegraphics*[width=2.1\columnwidth]{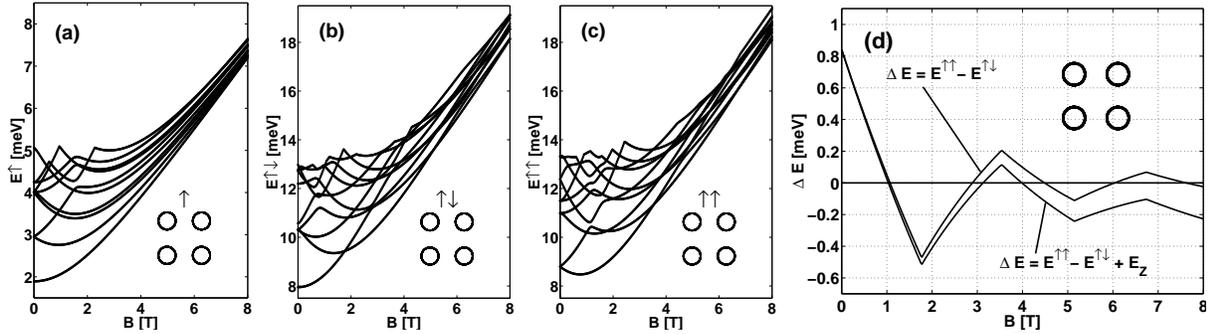}
\end{center}
\caption{Ten lowest energy levels of $L=10$ nm QDM for (a)
single-particle, (b) spin-singlet ($S=0$) and (c) spin-triplet ($S=1$)
states. Triplet-singlet energy difference of $L=10$ nm QDM is plotted
in (d).}
\label{Elev}
\end{figure*}

The energy differences of the lowest triplet and singlet states is
plotted in Fig. \ref{dE} as a function of magnetic field ($B$) at
different inter-dot spacings ($L=[0,20]$ nm). Ground state of
four-minima QDM consist of altering singlet ($S=0$) and triplet
($S=1$) states. At small magnetic field the ground state of a
two-electron QDM is spin-singlet ($S=0$), which changes to
spin-triplet ($S=1$) as the magnetic field is increased. The first
singlet-triplet transition can be be understood with simple occupation
of the lowest single-particle states: In the singlet state the two
electrons occupy the lowest energy eigenstate with opposite spins
($S=0$).  As the magnetic field increases, the energy difference of
the lowest and the second lowest single-particle levels decreases.  At
some point the exchange energy in the spin-triplet state becomes
larger than the energy difference between the adjacent energy
levels. Thus, the singlet-triplet transition occurs and the adjacent
eigenlevels are occupied with electrons of parallel spins ($S=1$).

However, the true solution of two-electron QDM is much more
complicated than the occupation of single-particle levels and the
inclusion of the exchange energy. Interaction between the electrons
changes the situation drastically. As a signature of complex many-body
features a second singlet state at higher magnetic field is observed
in double-minima QDM \cite{AriPRL}.

In four-minima QDM we find, as well, a second singlet region at higher
magnetic field, but also a third, fourth, and even a fifth singlet
state at the inter-dot spacings studied ($L \leq 20$ nm). Actually, it
is interesting to note that especially at large inter-dot spacings
rapid singlet-triplet oscillations are seen as a function of magnetic
field. However, when the Zeeman term (which lowers the spin-triplet
energy) is included in the energy, the subsequent singlet states after
the second $S=0$ are no longer ground states as can be seen in
Fig. \ref{dE} (b).

It is surprising to see how stable the second singlet is in the
four-minima QDM. Even if the Zeeman energy is included, there remains
a $0.7-1$ T magnetic field window of the second $S=0$  state at the
greatest studied distance. The energy difference of $L=10$ nm QDM can
be examined in Fig. \ref{Elev} (d). The third singlet region at $B
\approx [6,7.5]$ T is no longer ground state if the Zeeman term is
included as the lower curve indicates. However, the second singlet
persist as a ground state to the largest $L$ studied as
Fig. \ref{dE}~(b) indicates. This is in contrary to two-minima QDM
where the second singlet state is observed only at very small
inter-dot distances ($L \leq 2.5$ nm) if the Zeeman term is included
\cite{AriPRL}.

Ten lowest energy levels of $L=10$ nm QDM as a function of magnetic
field are shown in Fig. \ref{Elev} (a)-(c) for single-particle,
spin-singlet and spin-triplet states, respectively.

\section{Far-infrared spectra}

\begin{figure}
\begin{center}
\includegraphics*[width=\columnwidth]{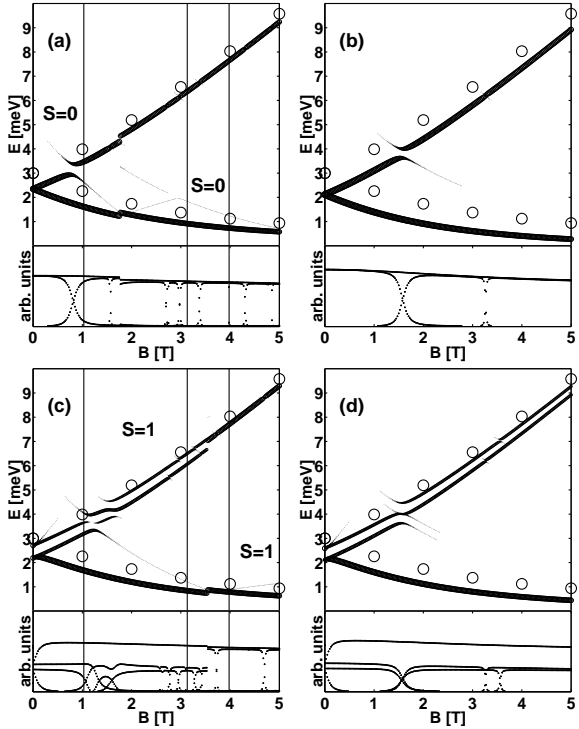}
\caption{Far-infrared spectra of $S=0$ in (a), $S=1$ in (c) and
corresponding non-interacting spectra spectra in (b) and (d) for $S=0$
and $S=1$, respectively.}
\label{FIR}
\end{center}
\end{figure}

The calculated FIR spectra of $L=10$ nm QDM are shown in
Fig.~\ref{FIR} for spin-singlet in (a) and spin-triplet in (c). For
the comparison we plot also non-interacting two-electron spectra for
the symmetric wave function (non-interacting $S=0$) and for the
antisymmetric wave function (non-interacting $S=1$) in Fig.~\ref{FIR}
(b) and (d), respectively. We like to mark that (b) is also
single-particle FIR spectra of $L=10$ nm four-minima QDM.

The linewidth at the corresponding energy (in meV) in FIR spectra
indicates the transition probability from the ground state to an
excited state and it is also plotted below each spectrum (in arbitrary
units). {\it E.g.} in $S=0$ spectrum (a) the upper $\omega_+$ mode has
rather constant transition probability till the first anticrossing,
where the transition probability of the lower part of the anticrossing
rapidly decreases as a function of $B$, whereas the upper part
increases at the same time. The vertical lines in (a) and (c) mark the
magnetic field values where the singlet-triplet or triplet-singlet
transition is observed. They also mark regions of observable spectra,
which is $S=0$ at magnetic field values $B \approx [0, 1]$ T and $B
\approx [3, 4]$ T, and $S=1$ outside these regions. In non-interacting
spectra no transitions of ground state occurs. The open circles
represent Kohn modes of parabolic QD with $\hbar\omega_0 = 3$ meV
confinement.
 
Anticrossings in the upper $\omega_+$ branch are seen in the singlet,
triplet and non-interacting spectra. These result in from the
non-circular confinement potential. Discontinuities in the interacting
spectra, at $B \approx 1.2$ T in $S=0$ and at $B \approx 3.5$ T in
$S=1$, can be identified to crossing ground states (see also
Fig. \ref{Elev} (b) and (c)). However, these crossings are observed in
the magnetic field values where the ground state is of the other
spin-type.  At higher magnetic field ($B>8$ T) one could see
discontinuities in the spin-polarized system ($S=1$) resulting from
the crossing ground state levels in the spin-triplet
spectra. Otherwise only discontinuities in the observable spectra
result from altering singlet and triplet FIR spectra.

As x- and y-excitations are degenerate, there occurs no zero-field
splittings of the Kohn modes which were observed in the two-minima QDM
\cite{MeriPRL}. In the spin-triplet spectra the upper branch is split
to two modes separated by a clear energy gap. Similar type of
splittings of the upper branch are seen also in other types of
non-circular potentials~\cite{MeriPRL,Ullrich}.  The comparison of
interacting and non-interacting spectra show remarkably similar
spectra. In non-interacting spectra anticrossings are in slightly
higher $B$ and modes are slightly lower in energy than the interacting
modes. As the Coulomb repulsion is present in the interacting case
electrons feel effectively steeper confinement resulting slightly
higher excitation energies.  Only notable difference is in the triplet
spectra, where the double structure of the $\omega_+$ changes to
single peak after $B \approx 3.5$ T in the interacting
case. Otherwise, interacting and non-interacting spectra are so
similar that we can conclude that the FIR spectra reflects mainly the
single-particle excitations of electrons in the low-symmetry
confinement and only minor changes are induced by the
electron-electron interactions.

\section{Summary}
To summarize, we have calculated the ground state of two electrons in
four-minima quantum-dot molecule as a function of magnetic field. Our
exact diagonalization calculations reveal a complicated two-body
ground state structures of altering singlet and triplet states as a
function of magnetic field.  In the far-infrared spectra we observe
anticrossings and an additional mode in the spin-triplet state both
arising from the non-circular confinement potential.  We conclude that
electron-electron interactions have only a minor effect on the FIR
spectra of four-minima QDM.

\end{document}